\documentclass[hyper,12pt,letterpaper]{JHEP3}
\usepackage{epic,eepic}
\usepackage{graphicx}

\title{Comments on the Spectrum of CHL Dyons}

\preprint{\hepth{0702150}\\TIFR/TH/06-32}
\author{Atish Dabholkar\footnote{Email:
atish@theory.tifr.res.in}$^{~1, ~2}$, Davide Gaiotto\footnote{Email:
dgaiotto@fas.harvard.edu}$^{~3}$, and Suresh Nampuri\footnote{Email:
suresh@theory.tifr.res.in}$^{~1}$\\
\it $^1$Department of Theoretical Physics\\
\it Tata Institute of Fundamental Research\\
\it Homi Bhabha Rd, Mumbai 400 005, India\\

\it $^2${Laboratoire de Physique Th\'eorique et Hautes Energies (LPTHE)\\
\it{Tour 24-25, 5\` eme \'etage, Boite 126, 4 Place Jussieu, 75252
Paris Cedex 05}}\\
\it{Unit\'e Mixte de Recherche (UMR 7589)}\\
\it{Universit\'e Pierre et Marie Curie-Paris 6; CNRS; Universit\'e
Denis Diderot-Paris 7}\\

\it $^3$Jefferson Labs,\\
\it 17 Oxford St, Harvard University\\
\it Cambridge, MA 02138  USA}

\abstract{We address a number of puzzles relating to the proposed
formulae for the degeneracies of dyons in orbifold compactifications
of the  heterotic string to four dimensions with $\CN =4$
supersymmetry. The partition function for these dyons is given in
terms of Siegel modular forms associated with genus-two Riemann
surfaces. We point out a subtlety in demonstrating S-duality
invariance of the resulting degeneracies and give a prescription
that makes the invariance manifest. We show, using M-theory lift of
string webs,  that the genus-two contribution captures the
degeneracy only if a specific irreducibility criterion is satisfied
by the charges. Otherwise, in general there can be additional
contributions from higher genus Riemann surfaces. We analyze the
negative discriminant states predicted by the formula. We show that
even though there are no big black holes in supergravity
corresponding to these states, there are multi-centered
particle-like configurations with subleading entropy in agreement
with the microscopic prediction and our prescription for S-duality
invariance. The existence of the states is moduli dependent and we
exhibit the curves of marginal stability and comment on its relation
to S-duality invariance.}

\newcommand{\CC}{\cal{C}}

\def\h{\eta}

\def\CN{{\cal N}}
\def\IC{\relax\hbox{$\inbar\kern-.3em{\rm C}$}}

\def\IC{{\bf C}}

\def\CN{{\cal N}}

\def\bea{\begin{eqnarray}}
\def\eea{\end{eqnarray}}
\def\be{\begin{equation}}
\def\ee{\end{equation}}
\def\ba{\begin{align}}
\def\ea{\end{align}}
\def\bse{\begin{subequations}}
\def\ese{\end{subequations}}
\def\1F1{{}_1\!F_1}
\def\2F0{{}_2\!F_0}

\def\h3{$\textrm{H}_3^+$}

\def\IC{{\mathbb C}}

\def\lbldef#1#2{\expandafter\gdef\csname #1\endcsname {#2}}

\def\href#1#2{#2}

\newcommand{\beq}{\begin{equation}}
\newcommand{\eeq}{\end{equation}}
\newcommand{\ber}{\begin{eqnarray}}
\newcommand{\eer}{\end{eqnarray}}

\def\be{\begin{eqnarray}}
\def\ee{\end{eqnarray}}

\begin{document}


\section{Introduction}

The spectrum of BPS states has played a very significant role in
several important developments in string theory. In the limit of
small charges, counting low-lying BPS states has provided many
nontrivial  tests of duality. In the opposite limit of large
charges, computation of asymptotic degeneracies of BPS states has
allowed for a statistical interpretation of the Bekenstein-Hawking
entropy of certain supersymmetric black holes addressing a
long-standing problem in quantum gravity. It is clearly desirable to
obtain similar quantitative information about the spectrum of BPS
states with arbitrary intermediate charges.

Many BPS states can be mapped by duality to either perturbative
winding-momentum states or a collection of  D-branes in a particular
duality frame. In both cases, using powerful techniques from the
worldsheet and from gauge theory, it is possible to obtain  detailed
exact information about the BPS spectrum for all values of charges.
To complete the picture,  one would like to develop tools to analyze
the spectrum of BPS states that are nonperturbative in all duality
frames. Such states, for example various dyons in four dimensions,
are in a sense more interesting because they can allow us to access
the interior regions of moduli space at intermediate couplings that
are not weakly coupled in any duality frame.

For heterotic string compactifications with $\CN =4$  supersymmetry
in four dimensions, there exists a formula for the exact
degeneracies of the dyonic quarter-BPS states in terms of Siegel
modular forms of $Sp(2, \mathbb{Z})$ and its
subgroups\cite{Dijkgraaf:1996it, Jatkar:2005bh}. As for the
perturbative winding-momentum states or for D-brane bound states, it
is desirable to have a systematic derivation of the dyon partition
function using worldsheet or gauge theory techniques. A weak
coupling derivation has been suggested recently using the 4d-5d lift
\cite{Gaiotto:2005gf, Shih:2005uc}. A similar derivation for the
more general orbifolds with ${\cal N}=4$ supersymmetry is discussed
in \cite{Dabholkar:2006xa,{David:2006ji},{David:2006yn},
{David:2006ud}}. The role of genus-two Riemann surfaces and the
appearance of $Sp(2, \mathbb{Z})$ can be explained using  an
M-theory lift of string webs \cite{Gaiotto:2005hc} and an
alternative derivation of the dyon partition function in terms of a
genus-two partition function of the heterotic string has been given
in \cite{Dabholkar:2006xa, Dabholkar:2006bj}.

Our purpose here is to address a number of subtleties in the
interpretation and derivation of the dyon partition function using
these ideas. We first summarize the basic ingredients of the dyon
partition function in $\S{\ref{Dyon}}$. In $\S{\ref{Sduality}}$ we
discuss the S-duality invariance  and give a prescription for the
choice of the contours that yields manifestly duality invariant
spectrum. In $\S{\ref{Higher}}$, using M-theory lift of string webs,
we show that the genus-two formula is adequate only if the charges
satisfy a specific irreducibility criterion. Otherwise there can be
additional contributions from higher genus Riemann surfaces. In
$\S{\ref{Negative}}$ we consider states predicted by the formula
that have negative discriminant for which there are no big black
holes in supergravity corresponding to these states. We discuss a
simple example for such states with negative discriminant in
$\S{\ref{Micro}}$, and show that there are multi-centered
particle-like configurations with subleading entropy in agreement
with the microscopic prediction as well as with our prescription for
S-duality invariance. The supergravity solutions analyzed in
$\S{\ref{Supergravity}}$ display an intricate moduli dependence. We
show in $\S{\ref{Marginal}}$ that a two-centered solution with the
desired degeneracy exists in a large region of moduli space where
supergravity and string loop expansion is under control. However,
there  a straight line in the axion dilaton space defines a line of
marginal stability. The state exists only on one side of the line,
decays into two fragments as one approaches this line, and ceases to
exist on the other side of the line.  We conclude in
$\S{\ref{Discussion}}$ with a discussion of the interpretation of
the dyon degeneracy formula in the light of above-mentioned
considerations of irreducibility criterion, moduli dependence, lines
of marginal stability, and duality invariance.

\section{The Dyon Partition Function \label{Dyon}}

Let $\Omega$ be a $(2\times 2)$ symmetric matrix with complex
entries
\begin{equation}\label{period}
   \Omega = \left(
              \begin{array}{cc}
                \rho & v \\
                v & \sigma \\
              \end{array}
            \right)
\end{equation}
satisfying
\begin{equation}\label{cond1}
   (\textrm{Im} \rho) > 0, \quad (\textrm{Im} \sigma) > 0, \quad
   (\textrm{Im} \rho)(\textrm{Im} \sigma) > (\textrm{Im} v)^2
\end{equation}
which parametrizes  the `Siegel upper half plane' in the space of
$(\rho, v, \sigma)$. It can be thought of as the  period matrix of a
genus two Riemann surface. For a genus-two Riemann surface, there is
a natural symplectic action of $Sp(2, \mathbb{Z})$ on the period
matrix. We write an element $g$ of $Sp(2, \mathbb{Z})$ as a
$(4\times 4)$ matrix in the block form as
\begin{equation}\label{sp}
   \left(
  \begin{array}{cc}
    A & B \\
    C & D \\
  \end{array}
\right),
\end{equation}
where $A, B, C, D$ are all $(2\times 2)$ matrices with integer
entries. They  satisfy
\begin{equation}\label{cond}
   AB^T=BA^T, \qquad  CD^T=DC^T, \qquad AD^T-BC^T=I\, ,
\end{equation}
so that $g^t J g =J$ where $J = \left(
                                  \begin{array}{cc}
                                    0 & -I \\
                                    I & 0 \\
                                  \end{array}
                                \right)$
is the symplectic form. The action of $g$  on the period matrix is
then  given by
\begin{equation}\label{trans}
    \Omega \to (A \Omega + B )(C\Omega + D ) ^{-1}.
\end{equation}
The object of our interest is a Siegel modular form $\Phi_k(\Omega)$
of weight $k$ which transforms as
\begin{equation}\label{phi}
    \Phi_k [(A \Omega + B )(C\Omega + D ) ^{-1}] =  \{\det{(C\Omega + D )}\}^k
    \Phi_k (\Omega),
\end{equation}
under an appropriate congruence subgroup of $Sp(2, \mathbb{Z})$
\cite{Jatkar:2005bh}. The subgroup as well as the  index $k$ of the
modular form are determined  in terms of the order $N$ of the
particular CHL $\mathbb{Z}_N$ orbifold one is considering
\cite{Jatkar:2005bh}. In a given CHL model, the inverse of the
$\Phi_k$ is to be interpreted then as a partition function of dyons.

To see in more detail how the dyon degeneracies are defined in terms
of the partition function, let us consider for concreteness the
simplest model of toroidally compactified heterotic string as in the
original proposal of Dijkgraaf, Verlinde, Verlinde
\cite{Dijkgraaf:1996it}. Many of the considerations extend easily to
the more general orbifolds with ${\cal N}=4$ supersymmetry. In this
case the relevant modular form is the well-known  Igusa cusp form
$\Phi_{10}(\Omega)$ of weight ten of the full group $Sp(2,
\mathbb{Z})$. A dyonic state is specified by the charge vector $Q=
(Q_e, Q_m)$ which transforms as a doublet of the S-duality group
$SL(2, \mathbb{R})$ and as a vector of the T-duality group $O(22, 6;
\mathbb{Z})$. There are three T-duality invariant quadratic
combinations  $Q_m^2$, $Q_e^2$, and $Q_e \cdot Q_m$ that one can
construct from these charges. Given these three combinations, the
degeneracy $d(Q)$ of dyonic states of charge $Q$ is then given by
\begin{equation}\label{degen}
    d(Q)=g\left({1\over 2}Q_m^2,\, {1\over 2}
     Q_e^2,\, Q_e\cdot Q_m\right)\, ,
\end{equation}
where $g(m, n, l)$ are the Fourier coefficients of $1/\Phi_{10}$,
\begin{equation}\label{fourier}
    {1\over \Phi_{10}(\rho, \sigma, v)} =
   \sum_{m\ge -1,n\ge -1, l}
   e^{2 \pi i(m \rho
     + n \sigma + l v)} g(m,n,l)\, .
\end{equation}
The parameters $(\rho, \sigma, v)$ can be thought of as the chemical
potentials conjugate to the integers $\left({1\over 2}Q_m^2,\,
{1\over 2}
     Q_e^2,\, Q_e\cdot Q_m\right)$ respectively.
The degeneracy $d(Q)$ obtained this way satisfies a number of
physical consistency checks. For large charges, its logarithm agrees
with the Bekenstein-Hawking-Wald entropy of the corresponding black
holes to leading and the first subleading order
\cite{Dijkgraaf:1996it, LopesCardoso:2004xf, Jatkar:2005bh,
LopesCardoso:2006bg, Shih:2005he}. It is integral as expected for an
object that counts the number of states. It is formally S-duality
invariant \cite{Dijkgraaf:1996it, Jatkar:2005bh} but as we will see
in the next section the formal proof is not adequate. An appropriate
prescription is necessary as we explain in detail in the next
section which also allows for a nontrivial moduli dependence.

\section{S-Duality Invariance \label{Sduality}}

The first physical requirement on the degeneracy $d(Q)$ given by
(\ref{degen}) is that it should be invariant under the S-duality
group of the theory. For the simplest case of toroidal
compactification that we are considering, the S-duality group is
$SL(2, \mathbb{Z})$ and more generally for $\mathbb{Z}_N$ CHL
orbifolds its a congruence subgroup $\Gamma_1(N)$ of $SL(2,
\mathbb{Z})$. So, we would like to show for the $N=1$ example, that
the degeneracy (\ref{degen}) is invariant under an S-duality
transformation
\begin{equation}\label{stranform}
    Q_e \to    Q_e' = a    Q_e + b    Q_m,
   \qquad Q_m\to Q_m' = c    Q_e + d    Q_m\, ,
   \qquad \pmatrix{a & b\cr c & d}\in SL(2, \mathbb{Z})\, .
\end{equation}
A formal proof of S-duality following \cite{Dijkgraaf:1996it,
Jatkar:2005bh} proceeds as follows. Inverting the relation
(\ref{fourier}) we can write
\begin{equation}\label{inverse3}
   d(Q) = \int_{\CC}
   d^3 \Omega \, e^{-i\pi Q^{\prime t} \cdot
     \Omega \cdot Q}\,
   {1 \over  \Phi_{10}(\Omega)}
\end{equation}
where the integral is over the contours
\begin{equation}\label{contour}
0 < \rho \leq 1, \quad 0 < \sigma \leq 1, \quad 0 < v \leq 1
\end{equation}
along the real axes of the three coordinates $(\rho, \sigma, v)$.
This defines the integration curve $\CC$ as a 3-torus in the Siegel
upper half plane. Now we would like to show
\begin{equation}\label{prime}
    d(Q') = \int_{\CC}
   d^3 \Omega \, e^{-i\pi Q^{\prime t} \cdot
     \Omega \cdot Q'}\,
   {1 \over  \Phi_{10}(\Omega)} \,
\end{equation}
equals $d(Q)$. To do so, we define
\begin{equation}\label{int}
 \Omega'\equiv\pmatrix{ \rho' &  v'
    \cr  v' & \sigma'}
   = ( A \Omega +  B) ( C \Omega + D)^{-1},
\end{equation}
for
\begin{equation}\label{sdual2}
     \pmatrix{ A &  B\cr
      C &  D}= \pmatrix{a & -b & b & 0\cr -c & d & 0 & c\cr
     0 & 0 & d & c\cr 0 & 0 & b & a}\, \in  Sp(2, \mathbb{Z})\,.
\end{equation}
We can change the integration variable from $\Omega$ to $\Omega'$.
Using these transformation properties and the modular properties of
$\Phi_{10}$ we see that
\begin{eqnarray}
   d^3 \Omega' &=& d^3 \Omega \, ,\\
  \Phi_{10}( \Omega') &=& \Phi_{10}(\Omega)\, , \\
  Q^{\prime t} \cdot
     \Omega'\cdot Q' &=& Q^{t} \cdot
     \Omega \cdot Q
\end{eqnarray}
Moreover, the integration contour $\CC$ as defined in
(\ref{contour}) is invariant under the duality transformation on the
integration variables (\ref{int}). We therefore conclude
\begin{equation}\label{inverse2}
     d(Q') = \int_{\CC}
   d^3 \Omega' \, e^{-i\pi Q^{\prime T} \cdot
     \Omega' Q'}\,
   {1 \over  \Phi_{10}(\Omega')} = d(Q)\, .
\end{equation}
This formal proof is however not quite correct. The reason is that
the partition function $1/\Phi_{10}$ has a double pole at $v =0$
which lies on the integration contour $\CC$. Thus the integral in
(\ref{inverse3}) is not well-defined on the contour $\CC$ and  one
must give an appropriate prescription for the integration. The
non-invariance can also be seen explicitly from the Fourier
expansion. To illustrate the point, let us look at states with
\begin{equation}\label{illust}
    \frac{1}{2} Q_m^2 = - 1, \quad \frac{1}{2} Q_e^2 = - 1, \quad Q_e
    \cdot Q_m = N.
\end{equation}
Then according to  (\ref{degen}), the degeneracy of such states can
be read off from the coefficient of $y^N /qp$ in the Fourier
expansion (\ref{fourier}). {}From the product representation of
$\Phi_{10}$ given for example in \cite{Dijkgraaf:1996it}, we see
that we need to pick the term that goes as $p^{-1}q^{-1} y^N$ in the
expansion of
\begin{equation}\label{expansion}
   \frac{1}{qp (y^{\frac{1}{2}} -y^{-\frac{1}{2}})^2} =  \frac{y}{qp}\frac{1}{(1
   -y)^2} = \sum_{N=1}^\infty N  q^{-1} p^{-1} y^N
\end{equation}
which implies that
\begin{equation}\label{d1}
    d(-1, -1, N) = N.
\end{equation}

Let us now look at what is required for invariance under $SL(2,
\mathbb{Z})$ transformations. Consider, for example, the element
\begin{equation}\label{S}
   S = \left(
         \begin{array}{cc}
           0 & 1 \\
           -1 & 0\\
         \end{array}
       \right)
\end{equation}
of the S-duality group which  takes $(Q_e,\, Q_m)$ to $(Q_m, -Q_e)$.
Hence $(\frac{1}{2}Q_m^2,\, \frac{1}{2}Q_e^2,\, Q_e\cdot Q_m)$ goes
to $(\frac{1}{2}Q_e^2,\, \frac{1}{2}Q_m^2,\, -Q_e\cdot Q_m)$.
Invariance of the spectrum under this element of the S-duality group
would then predict $d(-1, -1, -N) = d(-1, -1, N) = N$. However, from
the expansion (\ref{expansion}) we see that there are no terms in
the Laurent expansion that go as $y^{-N}$ and hence an application
of the formulae (\ref{degen}) and (\ref{fourier}) would give $d(-1,
-1, -N) = 0$ in contradiction with the prediction from S-duality.

This apparent lack of S-duality invariance is easy to fix with a
more precise prescription. Note that the function $(y^{\frac{1}{2}}
-y^{-\frac{1}{2}})^{-2}$ has a $\mathbb{Z}_2$ symmetry  generated by
the element $S$ of the S-duality group that takes $y$ to $y^{-1}$.
Under this transformation the contour $|y| < 1$ is not left
invariant but instead gets mapped to the contour $|y|
> 1$. The new contour cannot be deformed to the original one without
crossing the pole at $y =1$ so if we are closing the contour around
$y=0$ then we need to take into account the contribution from this
pole at $y=1$.  Alternatively, it is convenient to close the contour
at $y^{-1} =0$ instead of $y=0$. Then we do not encounter any other
pole and because of the symmetry of the function $(y^{\frac{1}{2}}
-y^{-\frac{1}{2}})^{-2}$ under $y$ going to $y^{-1}$,  the Laurent
expansion around $y$ has the same coefficients as the Laurent
expansion around $y^{-1}$. We then get,
\begin{equation}\label{expansion2}
   \frac{1}{pq (y^{\frac{1}{2}} -y^{-\frac{1}{2}})^2} =  \frac{y^{-1}}{pq}
   \frac{1}{(1 -y^{-1})^2} = \sum_{N=1}^\infty N p^{-1} q^{-1} y^{-N}.
\end{equation}
If we now define $d(-1, -1, -N)$ as the coefficient of $qpy^{-N}$ in
the expansion (\ref{expansion2}) instead of in the expansion
(\ref{expansion}) then $ d(-1, -1, -N) = N = d(-1, -1, N)$
consistent with S-duality.

States with negative $N$ must exist if states with positive $N$
exist not only to satisfy S-duality invariance but also to satisfy
parity invariance. The ${\cal N} =4$ super Yang-Mills theory is
parity invariant. Under parity, our state with positive $N$ goes to
a state with negative $N$ and the asymptotic values $\chi$ of the
axion also flips sign at the same time. Hence if a state with $N$
positive exists at $\chi = \chi_0$ then a state with $N$ negative
must exist at $\chi = - \chi_0$. Thus, the naive expansion
(\ref{expansion}) would give an answer inconsistent with parity
invariance and one must use the prescription we have proposed, to
satisfy parity invariance. Note that even though S-duality and
parity both take the states $(-1, -1, N)$ to $(-1, -1, -N)$ they act
differently on the moduli fields.

In either case, the important point is that to extract the
degeneracies in an S-duality invariant way, we need to use different
contours for different charges. The function $1/\Phi_{10}$ has many
more poles in the $(\rho, \sigma, v)$ space at various divisors that
are the $Sp(2, \mathbb{Z})$ images of the pole at $y =1$ and in
going from one contour to the other these poles will contribute.
Instead of specifying contours, a more practical way to state the
prescription is to define the degeneracies $d(Q)$ by formulae
(\ref{degen}) and (\ref{fourier}) first for charges that belong to
the `fundamental cell' in the charge lattice satisfying the
condition $\frac{1}{2}Q_m^2 \geq -1$, $\frac{1}{2}Q_e^2 \geq -1$,
and $Q_e\cdot Q_m \geq 0$. For these charges $d(Q)$ can be
represented as a contour integral for a contour of integration
around $p = q = y=0$ that avoids all poles arising as images of $y
=1$. This can be achieved by allowing $(\rho, v, \sigma)$ to all
have a large positive imaginary part as noted also in
\cite{David:2006yn}. For other charges, the degeneracy is defined by
requiring invariance under $SL(2, \mathbb{Z})$. The degeneracies so
defined are manifestly S-duality invariant. This statement of
S-duality invariance might appear tautologous, but its consistency
depends on the highly nontrivial fact that an analytic function
defined by $\Phi_{10}(\rho, \sigma, v)$ exists that is $SL(2,
\mathbb{Z})$ invariant.  Its pole structure guarantees that one gets
the same answer independent of which way the contour is closed.

The choice of integration contour is possibly related to moduli
dependence of the spectrum. To see this let us understand in some
detail what precisely is required for S-duality. Given a state with
charge $Q$ that exists for the values of the moduli $\varphi$, the
statement of S-duality only requires that the degeneracy $d(Q)$ at
$\varphi$ be the same as the degeneracy $d(Q')$ at $\varphi'$ where
$Q'$ and $\varphi'$ are S-duality transforms of $Q$ and $\varphi$
respectively. In many cases, one can then invoke the BPS property to
assume that as we move around in the moduli space, barring phase
transitions, the spectrum can be analytically continued from
$\varphi'$ to $\varphi$ to deduce $d(Q') = d(Q)$ at $\varphi$. This
argument is known to work perfectly for half-BPS states in theories
with ${\cal N} = 4$ supersymmetric but with lower supersymmetry or
for quarter-BPS states in ${\cal N} = 4$, generically there can be
curves of marginal stability. In such cases, states with charges
$Q'$ may exist for moduli values $\varphi'$ but not for $\varphi$
and similarly states with charges $Q$ may exist for moduli values
$\varphi$ but not for $\varphi'$. Therefore, there are two
possibilities for extracting the dyon degeneracies.
\begin{itemize}
  \item There are no curves of marginal stability in the
  dilaton-axion moduli space. In this case if two charge
  configurations $Q$ and $Q'$ are related by S-duality, then
  $d(Q) = d(Q')$.
  \item There are curves of marginal stability in the
  dilaton-axion moduli space. In this case one can say at most
  that $d(Q)$ at $\varphi$ equals $d(Q')$ at $\varphi'$.
\end{itemize}
We will return in $\S{\ref{Discussion}}$ to a further discussion of
these possibilities in the present context after considering
explicit examples of moduli dependence and lines of marginal
stability in $\S{\ref{Negative}}$.

\section{Irreducibility Criterion and  Higher Genus Contributions \label{Higher}}

One way to derive the dyon partition function is to use the
representation of dyons as string webs wrapping the $\bf T^2$ factor
in Type-IIB on $\bf K3 \times T^2$. Using M-theory lift, the
partition function that counts the holomorphic fluctuations of this
web can be related to the genus-two partition function of the
left-moving heterotic string \cite{Gaiotto:2005hc,
Dabholkar:2006xa,Dabholkar:2006bj}. The appearance of genus-two is
thus explained by  the topology of the string web. Such a derivation
immediately raises the possibility of contribution from higher genus
Riemann surface because string webs with more complicated topology
are certainly possible.  In this section we address this question
and show that the genus-two partition function correctly captures
the dyon degeneracies if the charges satisfy certain irreducibility
criteria. Otherwise, there are higher genus  corrections to the
genus-two formula.

There are various derivations of the dyon degeneracy formula, but
often they compute the degeneracies for a specific subset of
charges, and then use duality invariance to extend the result to
generic charges. Such an application of duality invariance assumes
in particular  that under the duality group $SO(22,6,\mathbb{Z})$
the only invariants built out of charges would be $Q_e^2$, $Q_m^2$,
and $Q_e \cdot Q_m$. This assumption is incorrect. If two charges
are in the same orbit of the duality group, then obviously they have
the same value for these three invariants. However the converse is
not true. In general, for arithmetic groups, there can be discrete
invariants which cannot be written as invariants of the continuous
group.

An example of a non-trivial invariant that can be built out of two
integral charge vectors is $I = gcd(Q_e \wedge Q_m)$, \textit{i.e.},
the gcd of all bilinears $Q_e^i Q_m^j - Q_e^j Q_m^i$. Our goal is to
show that the genus-two dyon partition function correctly captures
the degeneracies if $I=1$. Note that half-BPS states have $I=0$ and
hence are naturally associated with a genus-one surface. If $I>1$,
then there are additional zero modes for the dyon under
consideration and it would be necessary to correctly take them into
account for counting the dyons.

\begin{figure}
\centering
\includegraphics[width=3.0in, height=2.3in]{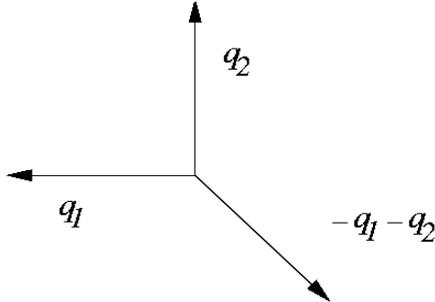}
\caption{Charge conservation at a string
junction}\label{fconservation}
\end{figure}

The  essential idea is to represent quarter-BPS states in the
Type-IIB frame as a periodic string network wrapped on the
two-torus. Type-IIB compactified on a $\bf K3$ has a variety of
half-BPS strings that can carry a generic set of $(21,5)$ charges
arising from D5 and NS5 branes wrapped on the $\bf K3$, D3-branes
wrapped on some of the $(19,3)$ two-cycles as well as D1 and
F1-strings. Several half-BPS strings can join into a web that
preserve a quarter of the supersymmetries \cite{{Schwarz:1996bh},
Aharony:1997bh,{Dasgupta:1997pu}, Sen:1997xi}. The supersymmetry
condition requires that the strings lie in a plane, and that their
central charge vectors also lie in a plane. The strands must be
oriented at relative angles that mimic the relative angle of their
central charge vectors. The condition on the angles between strings
guarantees the balance of tensions at the junction of three strands
of the web as shown in Fig.\ref{fconservation}.

\begin{figure}
\centering
\includegraphics[width=3.0in]{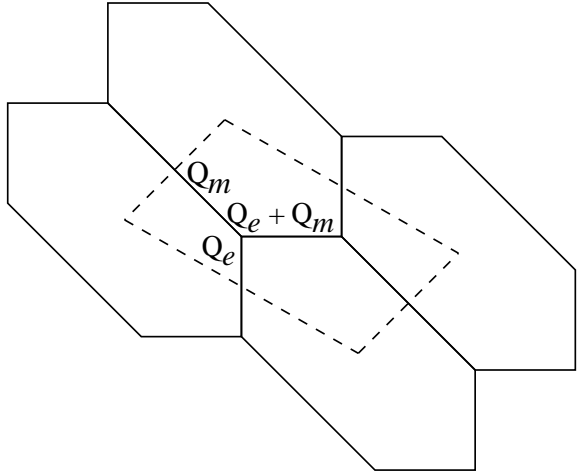}
\caption{A quarter-BPS dyon carrying irreducible charges $Q_e$ and
$Q_m$ with gcd($Q_e \wedge Q_m$)= 1}\label{web}
\end{figure}

The central charges are given in terms of the charges and the scalar
moduli of the theory as $Z = T q$. The matrix $T$ contains the
scalar moduli of the theory, that parameterize the way a vector in
the $\Gamma^{(21,5)}$ Narain lattice of charges decomposes into a
left-moving and a right-moving part. The five-dimensional
right-moving part is the vector of central charges for the string.
For generic values of the scalar moduli, one does not expect to have
tensionless strings. Hence it follows that $T Q = 0$ implies $Q =
0$. The condition that all central charges $T Q_i$ should lie in a
plane, $T Q_i = a_i T Q_1 + b_i T Q_2$ is then equivalent to $Q_i =
a_i Q_1 + b_i Q_2$, \textit{i.e.}, the charge vectors $Q_i$ of all
strings should also lie in a plane. A periodic string web can be
wrapped on the torus of a $\bf K3 \times T^2$ compactification as
shown in Fig.\ref{web}.

After compactification on the torus, the strands of the web can
carry additional charges: momentum along the direction they wrap,
and KK monopole charge for the circle they do not wrap. The charges
are organized in a $(22,6)$ charge vector. The result is a
quarter-BPS dyon in the four dimensional theory. A dyon with generic
charges $Q_e,Q_m$ typically has a very simple realization as a web
with three strands. A simple possible choice of charges on the
strands would be $Q_e$, $Q_m$, $Q_e + Q_m$.   This web comes from
the periodic identification of a hexagonal lattice. As the shape of
the $T^2$ or the moduli in $T$ change, the size of one strand may
become zero, and the web may degenerate into two cycles of the torus
meeting at a point. On the other side, of the transition the
intersection will open up in the opposite way and the configuration
then smoothly become a new three-strands web. For example, the web
with strands $Q_e$, $Q_m$, $Q_e + Q_m$ may go to a web with strands
$Q_e$, $Q_m$, $Q_e - Q_m$. This process has interesting consequences
on the stability of certain BPS states, that will be reviewed in
$\S{\ref{Negative}}$.

It has been argued \cite{Gaiotto:2005hc} that the partition function of
supersymmetric ground states for such webs can be computed by an
unconventional lift to M-theory that relates it to a chiral
genus-two partition function of the heterotic string. The genus-two
partition function computed using this lift for  CHL-orbifolds
\cite{Dabholkar:2006xa, {Dabholkar:2006bj}} indeed equals the dyon
partition function obtained by other means.

\begin{figure}
\centering
\includegraphics[width=3.1in]{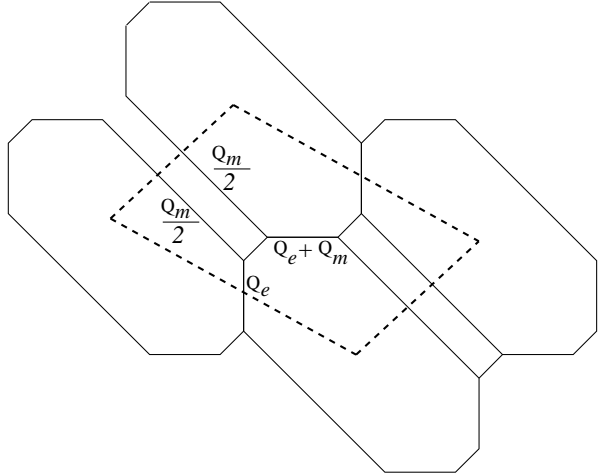}
\caption{A quarter-BPS dyon carrying reducible charges $Q_e$ and
$Q_m$ with gcd($Q_e \wedge Q_m$)=2}\label{split}
\end{figure}

A priori, the string junction need not to be stable against opening
up into more complicated configurations. For example, a strand may
split into two or more parallel strands, or the junction may open up
into a triangle. Any complicated periodic network made out of
strands with charges that are linear combinations of $Q_e$ and
$Q_m$, and such that the total charge flowing across one side of the
fundamental cell is $Q_e$, and through the other side $Q_m$ will
give a possible realization of the dyon as shown in Fig.\ref{split}.
If that is possible, the M-theory lift would predict a more
complicated expression for the dyon degeneracies. For simplicity, in
the following analysis we restrict to configurations with no
momentum or KK charge.

To understand the relation between the value of $I$ and the possible
variety of string webs that may describe a dyon with given charges,
it is useful to consider a graph in the space of charges that is
topologically dual to the string web. A dual graph is constructed as
follows. For every face of the web associate a vertex in the dual
graph. If two faces $A$ and $B$ in the web share an edge then the
corresponding vertices $A'$ and $B'$ in the dual graph are connected
by a vector that is equal in magnitude to the central charge of the
edge but rotated by $\pi/2$ in orientation compared to the edge.
Recall that each edge in the string web carries a central charge and
that the relative angles between the edges mimic the angles between
the corresponding central charge vectors. A junction has three faces
and three edges which maps to a triangle in the dual graph with
three vertices and three edges. Charge conservation at each junction
means that the vector sum of the three edge vectors is zero. This
then guarantees that the sides of the dual triangle actually close,
as their vector sum is zero. A string web constructed from a period
array of junctions then corresponds to a triangulation in the dual
graph.

Now, the vertices of the dual graph will sit at integral points of
the charge lattice, on the plane defined by the vectors $Q_e$ and
$Q_m$. The graph will have a fundamental cell with sides $Q_e$ and
$Q_m$. Our invariant $I$ counts the number of integral points inside
the fundamental cell. In this dual description, it is clear
geometrically that $Q_e^i Q_m^j - Q_e^j Q_m^i$ are the various
components of the area 2-form associated with the fundamental cell.
If all the components do not have common factor then the fundamental
parallelogram does not have any integral points either on the edges
or inside \cite{Verlinde:2003pv}.

Let us  see in more detail that $I$ counts the number of integral
points inside the fundamental cell. If all $Q_e^i Q_m^j - Q_e^j
Q_m^i$ are multiples of I, then consider any vector $Q$ such that $Q
\cdot Q_e$ is not a multiple of $I$. If such a vector does not
exist, then $Q_e$ is a multiple of $I$ and there are extra integral
points on the edges of the parallelogram. If on the other hand, such
a vector exists, then $\frac{Q \cdot Q_e}{I} Q_m^i - \frac{Q \cdot
Q_m}{I} Q_e^i$ is an integral charge vector that is a linear
combination of $Q_e$ and $Q_m$ with fractional coefficients. Up to
shifts by $Q_e$ and $Q_m$ it will lie inside the parallelogram.
Conversely, if the lattice of integral points that are coplanar with
$Q_e$ and $Q_m$ has a smaller fundamental cell than the
parallelogram with sides $Q_e$ and $Q_m$, then $Q_e = a Q_1 + b
Q_2$, $Q_m = c Q_1 + d Q_2$, $ad-bc >1$. There will be $ad-bc$
points inside the parallelogram, and as $Q_e \wedge Q_m = (ad-bc)
Q_1 \wedge Q_2$, $I = ad-bc$ is the number of points inside the
parallelogram with sides $Q_e$ and $Q_m$.

We thus see that if $I>1$, then the fundamental cell in the dual
graph has an internal integral point. Each of the internal points
can be used as a vertex for a triangulation. A generic periodic
triangulation subdivides a fundamental cell into at most $2I$ faces.
\footnote{This follows from Euler formula on the torus: a
triangulation has $3/2$ edges for each face, hence the number of
vertices is $1/2$ the number of faces.} In the dual description, a
string web on the torus that carries charges $Q_e$ and $Q_m$ will
have at most $2I$ three-strand junctions, and $I$ faces.

To put it differently, note that $I=1$ without any internal faces
is a genus two surface after M-theory lift. Adding a face increases
the genus by one. Hence the genus of a M-theory lift of a string web
with the invariant $I$ will be a surface with genus $I+1$.

When a face opens up at a string junction, its size is a zero mode
in that the mass of dyon is independent of the size of the
additional face. These zero modes and the invariant $I$ have been
discussed earlier in a related context of quarter-BPS dyons in field
theory using their realization as string junctions going between a
collection D3-branes \cite{Bergman:1998gs}. In that context, the
zero mode is one of the collective coordinates that must be
quantized to determine the ground state wavefunction. Similar
comments might apply in our case. More work is need to obtain a
definite interpretation of the higher genus contribution.

\section{States with Negative Discriminant \label{Negative}}

An important test of the dyon degeneracy formula is that for large
charges, the logarithm  of the predicted degeneracy $\log{ d(Q)}$
matches with Bekenstein-Hawking entropy. To make this comparison,
for a given a BPS dyonic state with electric and magnetic charges
$(Q_e, Q_m)$, one would like to find a supersymmetric black hole
solution of the effective action with the same charges and mass and
then compute its entropy. It is useful to define the `discriminant'
$\Delta$ by
\begin{equation}\label{discriminant}
    \Delta(Q) = Q_e^2 Q_m^2 -(Q_e \cdot Q_m)^2.
\end{equation}
which is the unique quartic invariant of the full U-duality group
$SO(22, 6; \mathbb{Z}) \times SL(2,\mathbb{Z})$. For a black hole
with charges $(Q_e, Q_m)$, the attractor value of the horizon area
is proportional to the square root of the discriminant and the
entropy is  given by
\begin{equation}\label{entropy}
    S(Q) = \pi \sqrt{\Delta(Q)}\,\,.
\end{equation}
On the microscopic side also, the discriminant is a natural
quantity. It is useful to think of $SL(2, \mathbb{Z})$  as $SO(1, 2;
\mathbb{Z})$ which has a natural embedding into $Sp(2, \mathbb{Z})
\sim SO(2, 3; \mathbb{Z})$. The dyon degeneracy formula depends on
the T-duality invariant vector of $SO(1, 2; \mathbb{Z})$
\begin{equation}\label{vector2}
    \left(
      \begin{array}{c}
         {Q_m^2}/{{2}}\\
        {Q_e^2}/{{2}} \\
        Q_e \cdot Q_m \\
      \end{array}
    \right)
\end{equation}
The discriminant is the norm of this vector with the Lorentzian
metric
\begin{equation}\label{metric}
\left(
        \begin{array}{ccc}
          0 & 2 & 0 \\
          2 & 0 & 0 \\
          0 & 0 & -1 \\
        \end{array}
      \right).
\end{equation}
With this norm, for a given state  $(Q_m^2/2, \, Q_e^2/2,\, Q_e
\cdot Q_m)$,the vector (\ref{vector2}) is spacelike, lightlike, or
timelike depending on whether $\Delta$ is positive, zero, or
negative. We can accordingly refer to the state as spacelike,
lightlike, or timelike.

Clearly, to obtain a physically sensible, nonsingular,
supersymmetric, dyonic black hole solution in supergravity, it is
necessary that the discriminant defined in (\ref{discriminant}) is
positive and large so that the entropy defined in (\ref{entropy}) is
real. The vector in (\ref{vector2}) in this case is spacelike. This
fact seems to lead to the following puzzle regarding the dyon
degeneracy formula. The formula  predicts a large number of states
that can have vanishing or negative discriminant. Since there are no
big black holes in supergravity in that case, there does not appear
to be a supergravity realization of these states predicted dyon
degeneracy. This raises the following question. \emph{Do the
lightlike and timelike states predicted by the dyon degeneracy
formula actually exist in the spectrum and if so what is their
macroscopic realization?} It is important to address this question
to determine the range of applicability of the dyon degeneracy
formula.

\subsection{Microscopic Prediction \label{Micro}}

To start with, let us emphasize that the lightlike or timelike
states are not necessarily pathological even though there is no
supergravity solution corresponding to them. The simplest example of
a lightlike state is the half-BPS purely electric state in the
heterotic frame with winding $w$ along a circle and  momentum $n$
along the same circle \cite{Dabholkar:1989jt, {Dabholkar:1990yf}}.
For such a state, $Q_e^2 = 2 nw $ is nonzero but since it carries no
magnetic charge, both $Q_m^2$ and $Q_e \cdot Q_m$ are zero and hence
the discriminant is zero. The supergravity solution is singular but
higher derivative corrections generate a horizon with the correct
entropy \cite{{Sen:1995in},Dabholkar:2004yr,Dabholkar:2004dq}. We
would like to know if similarly there exist quarter-BPS states that
are timelike or lightlike in accordance with the  predictions of the
dyon degeneracy formula and what their supergravity realization is.

In general, it is not easy to extract closed form asymptotics from
the degeneracy formula in this regime when the discriminant is
negative or zero. But we have already encountered a simple example
of a timelike state in $\S{\ref{Sduality}}$. The states with
$\left(Q_m^2/2,\,
     Q_e^2/2,\, Q_e\cdot Q_m\right)$ equal to $(-1, -1, N)$ have
discriminant $1-N^2$ which can be arbitrarily negative and we have
determined the degeneracy of this state to be  $d(-1, -1, N) =N$. Do
such states exist in the physical spectrum, and if so what is their
supergravity realization that can explain the degeneracy?

It is easy to construct such a state from a collection of winding,
momentum, KK5, NS5 states in heterotic description. We choose a
convenient representative that makes the supergravity analysis in
the following section simpler. We consider heterotic string
compactified on $\bf  T^4 \times S^1 \times {\tilde S}^1$. Let the
winding and momentum around the circle $S^1$ be $w$ and $n$ and
around the circle $\bf {\tilde S}^1$ be $\tilde w$ and $\tilde n$.
Similarly, $K$ and $W$ are the KK-monopole and NS5-brane charges
associated with the circle $\bf S^1$ whereas ${\tilde K}$ and
${\tilde W}$ are the KK-monopole and NS5-brane charges associated
with the circle ${\tilde S}^1$. Note that the state with charge $W$
can be thought of as an NS5 brane wrapping along $\bf T^4 \times
{\tilde S}^1$ whereas the states with charges ${\tilde W}$ is
wrapping along $\bf  T^4 \times S^1$. While the state that
magnetically dual to $n$ is $K$ in terms of Dirac quantization
condition, the state that is S-dual to $n$ is $W$. Similar comment
holds for other states. With this notation, we then choose the
charges $\Gamma=(Q_e|Q_m) = (n, w; \tilde n, \tilde w| W, K; \tilde
W, \tilde K)$ to be
\begin{equation}\label{gamma}
    \Gamma=(1, -1; 0, N|0, 0; 1, -1).
\end{equation}
This state clearly has  $(Q_m^2/2, Q_e^2/2,
Q_e\cdot{Q_m})=(-1,-1,N)$. We will show in the appendix
$\S{\ref{Supergravity}}$ that the supergravity solution
corresponding to this state with the required degeneracy has two
centers instead of one. One center is purely electric with charge
vector
\begin{equation}\label{electric}
\Gamma_1=(1, -1; 0, N|0, 0; 0, 0),
\end{equation}
and the other purely magnetic with charge vector
\begin{equation}\label{magnetic}
\Gamma_2=(0,0; 0,0|0, 0; 1, -1),
\end{equation}
both separated by a distance $L$. The corresponding  supergravity
solution exists for charge configuration with a positive, nonzero
value for the distance $L$ both for positive and negative $N$ in a
large regions of the moduli space but not for all values of the
moduli. We discuss the explicit solution and as well as the moduli
dependence and lines of marginal stability in the next subsections.

It is easy to see that such a two-centered solution has the desired
degeneracy  in agreement with the prediction from the dyon partition
function. Each center individually contributes no entropy because
for example the electric center by itself has $Q_e^2/2 =-1$ and
hence carries no left-moving oscillations. However, because the
charges are not mutually local, there is a net angular momentum $j=
N/2$ in the electromagnetic field. For large $N$, the angular
momentum multiplet has $2j +1 $ or $N$ states in agreement with the
dyon degeneracy formula. We thus see that at least some of the
states with negative discriminant predicted by the dyon degeneracy
formula can be realized physically but as multi-centered
configurations.

\subsection{Supergravity Analysis
\label{Supergravity}}

For the supergravity analysis of the dyonic configurations, it is
convenient to use the ${\cal N} =2$ special geometry formalism.
Consider Type-II string compactified on a Calabi-Yau threefold with
Hodge numbers $(h^{1,1}, h^{2,1})$ which results in a ${\cal N} =2$
supergravity in four dimensions with $h^{1,1}$ vector multiplets and
$h^{2,1} +1$ hyper-multiplets. The hypermultiplets will not play any
role in our analysis. The vector multiplet moduli space is a special
K\"ahler geometry parameterized by $h^{1, 1}+1$ complex projective
coordinates $\{X^I\}$ with $I = 0,1, \ldots, h^{1,1}$ and $\{\lambda
X^I\} \sim \{X^I\}$. The low energy effective action of the vector
multiplets is completely summarized by a prepotential which is a
homogeneous function $F(X^I)$ of degree two,
\begin{equation}\label{homo}
   F (\lambda X^I) = \lambda^2 F(X^I).
\end{equation}
In particular, the  K\"ahler potential $\cal K$ is determined in
terms of the prepotential by
\begin{equation}\label{kahler}
   e^{-{\cal K}} = i (\bar X^I F_I - X^I \bar F_I),
\end{equation}
where
\begin{equation}\label{fi}
    F_I = \frac{\partial F}{\partial X^I}.
\end{equation}

In our case, since we have a special Calabi-Yau $\bf K3 \times T^2$,
we actually get ${\cal N} =4$ supersymmetry which has two additional
gravitini multiplets. With our charge assignment, the vector fields
in the gravitini multiplets are not excited and we can restrict our
attention to the ${\cal N} =2$ sector. In the heterotic frame, we
have excited electric and magnetic charges (\ref{gamma}) which
couple only to gauge fields associated with the $\bf T^2$ part and
to the metric and the dilaton-axion. As a result, the sigma model
corresponding to the black hole configuration in $\bf R^4$ is
completely factorized into the $\bf T^4$ conformal field theory and
the sigma model involving $\bf T^2 \times R^4$ parts. This implies
that for analyzing our charge configuration, we can restrict our
attention to the moduli fields associated with $\bf T^2$ and the
dilaton-axion. The prepotential in this case can be chosen to be
\begin{equation}\label{prepot}
F(X^I)=-\frac{X^1 X^2 X^3}{X^0},
\end{equation}
which corresponds to the so called STU model. Here
\begin{equation}
S=X^1/X^0= a +i e^{-2\Phi}
\end{equation}
is the dilaton-axion field, where $a$ is the axion and $\Phi$ is the
dilaton in the heterotic frame. Similarly $ T=X^2/X^0$ is the
complex structure modulus of the $\bf T^2$ and $ U=X^3/X^0 $ is the
K\"ahler modulus of the $\bf T^2$ in the heterotic frame. All other
moduli fields do not vary in the geometry corresponding to our
charge configuration. Restricting to the STU model greatly
simplifies  the analysis. Indeed this motivates the choice of the
charges as in (\ref{gamma}).

Given the prepotential (\ref{prepot}) specifying the special
geometry, there is a natural symplectic action $Sp(4, \mathbb{R})$
on $(X^I, F_I)$. Similarly, the charges $(p^I, q_I)$ transform as a
symplectic vector. These charges are more naturally defined in the
Type-IIA frame, where $q_I$ are the electric charges arising from
D0-brane and wrapped D2-branes, and $p^I$ are the magnetic charges
arising from D6-brane and wrapped D4-branes.

A general supersymmetric multi-centered dyonic solution has a metric
of the form
\begin{equation}\label{met}
-e^{-2G({\vec r})}(dt+ \omega_i {dx}^i)^2 + e^{2G({\vec
r})}(dr^2+r^2 d\Omega_{2}^2).
\end{equation}
The four complex moduli fields  $X^I$ that solve the equations of
motion are determined in terms of the function $G$ and harmonic
functions $H^I$ and $H_I$ by the eight real equations
\begin{eqnarray}\label{sol}
  e^{-G}(X^I-\bar{X^I}) &=& H^I(\vec r) \\
  e^{-G}(F_I-\bar{F_I}) &=& H_I(\vec r),
\end{eqnarray}
in the gauge
\begin{equation}\label{gauge}
    e^{-\cal K} = \frac{1}{2}
\end{equation}
with the K\"ahler potential given by (\ref{kahler}). For a
configuration with $s$ charge centers with charges $\Gamma_a =
(p^I_a, q_{Ia}), a = 1, \ldots s$ localized at the centers $\vec r =
\vec r_a$ respectively, the  harmonic functions $H_I$ and $H^I$ are
given by \cite{Bates:2003vx}
\begin{equation}\label{harmo}
H^I=h^I + \sum_{a=1}^s\frac{p^I_a}{|{\vec r}-{\vec r}_a|}, \qquad
H_I=h_I + \sum_{a=1}^s \frac{q_{Ia}}{|{\vec r}-{\vec r}_a|}.
\end{equation}
The constants of integration $h_I$ and $h^I$ will  be determined in
terms of the moduli fields shortly. Let $\Sigma(Q)$ be the entropy
of the black hole which in our case equals $\pi \sqrt{\Delta(Q)} $.
Then geometry of the solution is completely determined in terms of
these harmonic functions \cite{Bates:2003vx}. The moduli are given
by
\begin{equation}
\frac{X^A}{X^0}=\frac{\partial_A \Sigma(H)-i H^A}{\partial_0
\Sigma(H) - i H^0}
\end{equation}
with $A = 1, 2, 3$  and $\partial_{A} = \partial
/{\partial{H^{A}}}$. The metric is given by
\begin{equation}
e^{-2G}=\Sigma(H),
\end{equation}
\begin{equation}
\nabla \times \omega = H^I \nabla H_I - H_I \nabla H^I.
\end{equation}
Taking divergence of both sides then implies the Denef's constraint
\cite{Denef:2000nb}
\begin{equation}
H^I {\nabla}^2 H_I - H_I {\nabla}^2 H^I = 0.
\end{equation}
This is a consistency condition for  a solution with $s$ centers to
exist, where $\nabla^2$ is the flat space Laplacian in $\bf R^3$.
This implies the following $s$ equations
\begin{equation}\label{denef}
(h_I  p^I_a -h^I  q_{Ia}) +\sum_{b=1}^s \frac{(p^I_a q_{Ib}- q_{Ia}
p^I_b)}{|\vec r_a- \vec r_b|}  = 0,
\end{equation}
where sum over repeated $I$ index is assumed. Summing over the index
$a$ in the equation above gives the summed constraint
\begin{equation}\label{sumcontraint}
 (h_I {p^I} - h^I q_I) = 0,
\end{equation}
where $p^I = \sum p^I_a$ and $q_I = \sum q_{Ia}$ are the total
charges.

The  values of the moduli fields $S=S_1+i S_2$, $T=T_1+i T_2$ and
$U=U_1+i U_2$ at asymptotic infinity are specified by six real
constants. The solutions on the other hand are determined by eight
real constants of integration $(h^I, h_I), I =0, 1, 2, 3$ which
however must satisfy two real constraints (\ref{gauge}) and
(\ref{sumcontraint}). Thus, they can be determined in terms of the
six asymptotic values of the moduli fields and the complete
supersymmetric solution for all fields is then determined by
(\ref{sol}), (\ref{met}), and (\ref{harmo}).

Specializing to our case, we will consider a two-centered solution
so $s=1,2$. We restrict ourselves to a region of moduli space where
$\bf T^2$ is factorized into two circles $\bf S^1 \times \tilde S^1$
and there is no $B$ field on the torus. In other words, we work on
the submanifold of the moduli space with $T_1 = U_1 =0$. Let $R_1$
and $R_2$ be the radii of the circles $\bf S^1$ and $\bf \tilde S^1$
respectively, $\chi$ be the asymptotic  expectation value of the
axion, and ${g^2}$ be the string coupling given by the asymptotic
value of $e^{2 \Phi}$. A nonzero value of $\chi$ will be essential
to obtain a well defined solution. Given this asymptotic data
\begin{equation}\label{asym}
    S_\infty=\chi+\frac{i}{g^2}, \qquad
T_\infty=i\frac{R_1}{R_2}, \qquad  U_\infty=i R_1 R_2,
\end{equation}
we now proceed to determine the constants of integration $(h^I,
h_I)$.

At asymptotic  infinity, $G(\vec r)$ vanishes, so the solutions
(\ref{sol}) reduce to
\begin{equation}
2 \textrm{Im}(X^I)=h^I, \qquad 2 \textrm{Im} (F_I)=h_I.
\end{equation}
Let $X^0_\infty=\alpha+ i \beta$.  Then from (\ref{sol}) and
(\ref{asym}) we see that the constants of integration are given by
\begin{eqnarray}\label{const}
  h^0 &=& 2\beta \qquad\qquad\qquad\qquad\quad h_0 =
  -2\alpha\frac{R_1^2}{g^2}-2 R_1^2
  \beta \chi\\
  h^1 &=& 2\alpha\frac{1}{g^2}+ 2 \beta \chi
  \qquad\qquad\,\,\,\,\,
h_1 = 2 \beta R_1^2 \\
h^2 &=& 2\alpha\frac{R_1}{R_2} \qquad\qquad\qquad\quad\,\,\,
h_2=2\beta
\frac{R_1 R_2}{g^2}-2\alpha\chi R_1 R_2\\
h^3 &=& 2\alpha R_1R_2 \qquad\qquad\qquad\,\,\, h_3=2\beta
\frac{R_1}{R_2 g^2}-2\alpha\chi \frac{R_1}{R_2}\, .
\end{eqnarray}
The two constants $\alpha$ and $\beta$ that we have introduced are
in turn determined in terms of the charges by plugging (\ref{const})
into the two constraint equations (\ref{gauge}) and
(\ref{sumcontraint}). Equation (\ref{gauge}) in particular implies
\begin{equation}\label{x0}
    |X^0|^2 =  \alpha^2+\beta^2 = \frac{1}
    {16 S_2 T_2 U_2}=\frac{g^2}{16 R_1^2}.
\end{equation}

\subsection{Moduli Dependence and Lines of Marginal Stability
\label{Marginal}}

So far our analysis is valid for any charge assignment but with the
specific choice of the asymptotic moduli as in (\ref{asym}). The
remaining equations (\ref{sumcontraint}) as well as (\ref{const})
depend on the specific charge assignment of the configuration under
study. To use the attractor equations to analyze the geometry for
our charge configuration (\ref{electric}) and (\ref{magnetic}), we
first translate the charges given in the heterotic frame to purely
D-brane charges in the Type-IIA frame. The charges $(p^I, q_I)$ in
the Type-IIA arise from various D-branes wrapping even-cycles. We
label charges so that $q_0$ is the number of D0-branes, $q_1$ is the
number of D2-branes wrapping the $\bf T^2$, $q_2$ is the number of
D2-branes wrapping a 2-cycle $\Sigma_2$ in $\bf K3$, $q_3$ is the
number of D2-branes wrapping a 2-cycle $\tilde \Sigma_2$ that has
intersection number one with the cycle $\Sigma_2$. Similarly, $p^0$
is the number of D6-branes wrapping $\bf K3 \times T^2$, $p^1, p^2,
p^3$ are the number of D4-branes wrapping $\bf K3$, $\tilde \Sigma_2
\times \bf T^2$ and $\Sigma_2 \times \bf T^2$ respectively. By the
duality chain in appendix B, these charges in the Type-IIA frame are
related to the electric and magnetic charges $(Q_e, Q_m)$ in the
heterotic frame by
\begin{equation}\label{elec}
Q_e =(n, w; \tilde{n}, \tilde{w})\equiv (q_0, -p^1, q_2, q_3)
\end{equation}
\begin{equation}\label{mag}
Q_m =(W, K; \tilde{W}, \tilde{K})\equiv (q_1, p^0, p^3, p^2)\,.
\end{equation}
Now we are ready to apply the ${\cal N} =2$ formalism to our
two-centered configuration with the charge assignment
(\ref{electric}) and (\ref{magnetic}). The electric center has
charges
\begin{equation}\Gamma_1 = (1,-1,0,N|,0,0,0,0)
\end{equation}
and the magnetic center has charges
\begin{equation}
\Gamma_2 = (0,0,0,0|0,0,1,-1).
\end{equation}
The constraint (\ref{sumcontraint}) then reads
\begin{equation} h_1-h^0-N h^3 + h_3-h_2 =0\, .
\end{equation}
Substituting the values of the integration constants $h^I$ and $h_I$
in terms of $\alpha$ and $\beta$ from (\ref{const}) into this
equation, we obtain one equation for the two unknowns $\alpha$ and
$\beta$ in terms of charges and asymptotic moduli
\begin{equation}
\beta (R_1^2-1+\frac{R_1}{R_2 g^2}(1-R_2^2))+\alpha (-N R_1 R_2-
\chi \frac{R_1}{R_2} +\chi R_1 R_2)=0
\end{equation}
Combining this with  the second equation (\ref{x0}) that comes from
the gauge fixing constraint $e^{-{\cal K}} =\frac{1}{2}$
(\ref{gauge}), we can now solve for the two unknowns to obtain
\begin{equation}
\alpha =\frac{(R_1^2-1+\frac{R_1}{R_2 g^2}(1-R_2^2))g}{4 R_1 (N R_1
R_2+ \chi \frac{R_1}{R_2}-\chi R_1 R_2)\Lambda}, \qquad \beta =
\frac{g}{4 R_1 \Lambda}\, ,
\end{equation}
where
\begin{equation}
\Lambda^2 = {1+ \left(\frac{R_1^2-1 + \frac{R_1}{R_2
g^2}(1-R_2^2)}{-N R_1 R_2- \chi \frac{R_1}{R_2} + \chi R_1
R_2}\right)^2}\,\,\, .
\end{equation}
We have thus determined the integrations constants (\ref{const})
that appear in the solution (\ref{harmo}) completely in terms of the
asymptotic moduli and the charges. The geometry of the solution is
in tern determined entirely in terms of the harmonic functions. In
particular the separation $L$ between the two centers can be
obtained by solving Denef's constraint (\ref{denef}), which for our
configuration becomes
\begin{equation}
h_2 - h_3 = \frac{N}{L}
\end{equation}
 we have,
\begin{equation}\label{distance}
2\frac{R_1}{R_2}({R_2}^2 -1 )(\frac{\beta}{g^2}-\alpha
\chi)=\frac{N}{L}
\end{equation}
Since $L$ is the separation between the two centers, it must be
positive. This requires that $(\frac{\beta}{g^2}-\alpha \chi)$ must
be  positive. It is clear that this can be ensured for a large
region of moduli space. The locus in the moduli space where this
quantity becomes negative determines the line of marginal stability
in the upper half $S$ plane by the equation
\begin{equation}\label{ineq}
\frac{1}{g^2}-  (\frac{R_1^2-1+\frac{R_1}{R_2 g^2}(1-R_2^2)}{N R_1
R_2+ \chi \frac{R_1}{R_2}-\chi R_1 R_2}) \chi = 0,
\end{equation}\label{line}
which simplifies to
\begin{equation}\label{ineq2}
\chi = N \frac{R_1 R_2}{R_1^2 -1} \frac{1}{g^2}.
\end{equation}
This equation defines a straight line in the complex $S_\infty$
plane with $S_\infty = \chi + i/g^2$. Note that the slope of the
line is proportional to $N$. For fixed $R_1$ and $R_2$, this defines
a curve of marginal stability in the complex $S_\infty$ plane.  For
positive $N$, the desired two-centered solution exists  if $\chi +
i/g^2$ lies to the left of the line defined by the equation
(\ref{ineq2}). In this region, the distance between the two centers
determined by Denef's constraint (\ref{distance}) is positive and
finite. After crossing the line of marginal stability, the solution
ceases to exist because then there is no solution with positive $L$
to the constraint (\ref{distance}). As one approaches the line of
marginal stability from the left, the distance $L$ between the
electric and magnetic centers  goes to infinity. In other words, the
total state with charge vector $\Gamma$ decays into two fragments
with charge vectors $\Gamma_1$ and $\Gamma_2$. The mass $M$ of the
state with charge $\Gamma$ is given in terms of the central charge
by the BPS formula $M = |Z(\Gamma)|$ with
\begin{equation}\label{central}
    Z = e^{{\cal K} /2} (p^I F_I - q_I X^I).
\end{equation}
At the curve of marginal stability, it is easy to check that
$Z(\Gamma) = Z(\Gamma_1) + Z(\Gamma_2)$. Hence the state with charge
vector $\Gamma$ can decay into its fragments with charge vectors
$\Gamma_1$ and $\Gamma_2$ by a process that is marginally allowed by
the energetics and charge conservation.

Similarly, if $N$ is negative, the straight line defined by
(\ref{line}) has negative slope and a solution with positive $L$
exists only to the right of this line. As we have noted, the
S-transformation maps the configuration with $N$ positive to $N$
negative. Hence the line with positive slope gets mapped to a line
with negative slope and thus the curves of marginal stability move
under S-duality. The fact that a two centered solution exists for
both signs and with the correct degeneracy is consistent with our
prescription for extracting S-duality invariant spectrum proposed in
$\S{\ref{Sduality}}$. In the wedge between the two lines defined the
two lines of marginal stability for $N$ positive and $N$ negative,
both states coexist. In other regions, only one or the other state
exists.

The simplicity of the line of marginal stability defined by
(\ref{ineq2}) has a simple and beautiful interpretation from the
string web picture reviewed in $\S{\ref{Higher}}$. Indeed a string
web made out of strands with certain charges exists only if these
charges can be carried by a supersymmetric string in six dimensions.
If one crosses a line of degeneration  in the moduli space, across
which a strand with charges, say, $Q_e + Q_m$ shrinks to zero length
and is replaced by a strand with charge $Q_e - Q_m$, the quarter-BPS
state will decay if no supersymmetric string with charge $Q_e - Q_m$
exists. The line of degeneration is simply the line at which a
string of charge $Q_e$ along one cycle of the torus and a string of
charge $Q_m$ along the other can be simultaneously supersymmetric.
This is equivalent to the requirement that the phase of $S$ is the
same as the angle between the central charge vectors for $Q_e$ and
$Q_m$, that defines a straight line in the $S$ plane. In the present
case $Q_e = (1,-1,0,N)$ and $Q_m = (0,0,1,-1)$, hence $Q_e \pm Q_m =
(1,-1,\pm 1,N\mp1)$. $\frac{1}{2}(Q_e \pm Q_m)^2 =-1 \pm N$, but a
BPS string with charge $Q$ must have ${Q^2}/{2}\geq-1$. Hence the
line of degeneration of the string web is indeed a line of marginal
stability.

It is not surprising that the existence of quarter-BPS dyons depends
on the moduli and that there are lines of marginal stability which
separate the regions where the state exists from where it does not
exist. This phenomenon is well known in the field theory context
\cite{Bergman:1997yw}. Moduli dependence of the spectrum of
quarter-BPS dyons and the lines of marginal stability in the present
string-theoretic context have been observed and analyzed from a
different perspective also in the forthcoming publication
\cite{Sen:2007nn}.

\section{Interpretation and Discussion\label{Discussion}}

As we have seen, the interpretation of the proposed dyon degeneracy
formula presents many subtleties. It is unlikely that the formula is
valid in all regions of moduli space for all charges in a way
envisioned in \cite{Dijkgraaf:1996it, {Jatkar:2005bh}} that depends
only on the three invariants $Q_e^2/2$, $Q_m^2/2$, and $Q_e\cdot
Q_m$. We summarize below our observations and what we believe would
be the consistent physical interpretation of the dyon degeneracy
formula.
\begin{itemize}
  \item It is clear that the  three invariants $(Q_e^2/2, Q_m^2/2, Q_e \cdot Q_m)$
do not uniquely specify the state and the degeneracy will depend on
additional data. This is natural because the arithmetic duality
group has many more invariants than the continuous duality group. We
have identified a particular invariant $I$ which determines when the
genus-two partition function is adequate but this is not the end of
the story. To illustrate this point, let us consider an even more
striking example of a quarter-BPS lightlike state for which
additional data is required to specify the degeneracy of
states.\footnote{We thank Boris Pioline for discussions on this
point.} Consider a perturbative BPS state that is purely electric in
the Type-IIA frame carrying winding $w$ along a circle of the $\bf
T^2$ factor and momentum $n$ along the same circle. In the heterotic
frame it corresponds to a state with $w$ NS5-branes wrapping $\bf
T^4 \times S^1$ with momentum $n$ along the $\bf S^1$. For nonzero
$n$ and $w$ the state carries arbitrary left-moving oscillations
$N_L = nw$ and has entropy $2 \pi \sqrt{2} \sqrt{nw}$. Unlike a
similar heterotic electric state which is half-BPS, these states are
quarter-BPS because both right and left movers carry supersymmetry
for the Type-II string. Now, for all such states, all three
invariants $(Q_e^2/2, Q_m^2/2, Q_e \cdot Q_m)$ vanish and so does
the discriminant. Thus there is a large set of legitimate
quarter-BPS states with the same values for the three invariant,
namely zero, but very different entropy depending on the values of
$n$ and $w$. The degeneracy of such states cannot possibly be
captured by the genus-two partition function. This example
illustrates that additional data might be required to determine the
degeneracy of states, although alternative explanations are
possible. The difference might also be attributed to a difference
between the absolute degeneracy of states and the supersymmetry
index computed by the dyon degeneracy formula.

  \item The states with negative discriminant appear problematic
at first because there is no black hole corresponding to them. We
have seen that they can nevertheless have a sensible physical
realization. In the specific example considered here the states are
described as a two-centered configuration in supergravity. These
configuration have the right degeneracy coming from the angular
momentum multiplicity consistent with the prediction of the dyon
degeneracy formula. We would like to propose that  other negative
discriminant states also exist and can be realized as complicated
multi-centered configurations. The supergravity analysis also
indicates that existence of these states is moduli dependent. The
states exist over a large region of the moduli space but cannot
exist in certain regions of the moduli space because the distance
between the two centers determined by Denef's constraint goes to
infinity. This shows that generically there are walls of marginal
stability in the moduli space that separate regions where the states
exist from regions where they do not. This is not surprising since
even in field theory, quarter-BPS states in ${\cal N} =4$ theories
are known to have curves of marginal stability \cite{Bergman:1997yw,
Argyres:2000xs}. It is possible that this moduli dependence is
related to the need to change the choice of contour to obtain an
S-duality invariant answer. As these lines of marginal stability
have a simple description in the string web picture, it might be
possible to understand the change of contour from the M-theory lift
of the string web.

\item Despite these subtleties, it is also true that the dyon partition function has been
derived from various points of views for specific charge
configurations and in specific regions of moduli space. Considering
the caveats above, a conservative interpretation of these results in
our view is that the dyon degeneracy formula given in terms of the
genus-two Siegel modular forms is exact and valid for specific
charges in the specific regions of moduli spaces as well as for all
charges related by a duality transformations in the dual regions of
the moduli space. This already contains highly nontrivial
information about the degeneracies of quarter-BPS bound states of
various branes in the theory. This can be seen quite generally from
the point of view of the string web picture. For a given charge
configuration, and in a given region of the moduli space, if a
string web is stable and can be lifted to a wrapped $\bf K3$-wrapped
M5-brane with a genus-two world sheet, then one can derive the
degeneracy from the genus-two partition function of the left-moving
heterotic string as has been done in \cite{Gaiotto:2005hc,
{Dabholkar:2006xa}, Dabholkar:2006bj}. However, as one moves around
in the moduli space, the string web can become unstable. Once the
string web is unstable, the dyon degeneracies can no longer be
obtained from the genus-two partition function. Thus the derivation
of the dyon partition function is valid in only a certain region of
the moduli space for a given charge configuration. Moreover, for
some quarter-BPS state, it may not be possible at all to represent
the state as a string web that lifts to a $\bf K3$-wrapped M5-brane.
For example, the Type-II perturbative states discussed above lift to
a circle-wrapped M2-brane with genus-one topology and not to a $\bf
K3$-wrapped M5-brane with genus-two topology. A circle-wrapped
M2-brane is nothing but the Type-II string and hence for these
states the counting is correctly done using the genus-one partition
function of the Type-II string and not using a genus-two partition
function of the heterotic string. These examples clearly delineate
the range of applicability of the dyon degeneracy formula.
\end{itemize}

\subsection*{Acknowledgements}

It is a pleasure to thank  Oren Bergman, K. Narayan, Boris Pioline,
Greg Moore, and Ashoke Sen for valuable discussions. A.~D. would
like to acknowledge the hospitality of the Aspen Center for Physics
and the theory group at the ASICTP where part of this work was
completed.

\appendix

\bibliographystyle{JHEP}

\bibliography{comments}

\end{document}